\documentstyle[aps,preprint]{revtex}
\begin{document}
\title{Exact Solution of a Hubbard Chain with \\
Bond-Charge Interaction}
\author{A.A. Aligia and L. Arrachea\thanks{Permanent address Departamento
de F\'{\i}sica, Universidad Nacional de La Plata, 1900 La Plata, Argentina}}
\address{Centro At\'{o}mico Bariloche and Instituto Balseiro \\
Comisi\'on Nacional de Energ\'{\i}a At\'{o}mica\\
8400 Bariloche, Argentina}
\maketitle
\begin{abstract}
\par We obtain the exact solution of a general Hubbard chain with
kinetic energy $t$, bond-charge interaction $X$ and on-site
interaction $U$ with the only restriction $t = X$. At zero
temperature and half filling, the model exhibits a Mott transition
at $U = 4t$. In the metallic phase and near half filling,
superconducting states are part of the degenerate ground state and
are favored for small $U$ if the system is slightly perturbed.
\end{abstract}
\newpage
\par The exact solutions, particularly those obtained using the
Bethe ansatz, have brought a very important progress in the
understanding of strongly correlated systems. However the
conditions for integrability using the Bethe ansatz are very
restrictive and only a limited class of realistic models can be
solved with this technique \cite{see}. Due to the importance of the
exact solutions in clarifying the effect of different physical
ingredients and as a test of approximations, the search of exact
solutions has been recently extended to other models and
techniques, in spite of the fact that in some cases the model or
the parameters are rather unrealistic \cite{mon,geb,es1,es2,bra,s1,s2}.
\par The model we consider is a particular case of the following Hamiltonian:
\begin{eqnarray}
H = H_U + H_t = U \sum_i n_{i \uparrow} n_{i \downarrow} +
\nonumber \\
\sum_{<ij> \sigma} c^\dagger_{j - \sigma} c_{i - \sigma} \left \{ t_{AA} (1 -
n_{i \sigma}) (1 - n_{j \sigma}) + \right. \nonumber \\
\left. t_{AB} [n_{i \sigma} (1 - n_{j \sigma}) + (1 - n_{i \sigma}) n_{j
\sigma} ] + t_{BB} n_{i \sigma} n_{j \sigma} \right \} .
\end{eqnarray}
$H$ has been derived as an effective one-band Hamiltonian for
the description of cuprate superconductors \cite{sim}. Similar
models including in some cases the nearest-neighbor repulsion $V$
have been studied by several authors
\cite{es1,es2,s2,sim,hir,sev,bar,jap}. If $t_{AA} + t_{BB} -
2t_{AB} = 0$ the three-body term of $H_t$ vanishes, and $H$ reduces
to the model considered by Hirsch and Marsiglio, in the framework of their
theory of ``hole
superconductivity'' \cite{hir}. Following Ref. \cite{s2}, we
call the coefficients of the one-and two-body parts of $H_t$ as
$t_{AA} = -t$ and $t_{AB} - t_{AA} = X$ respectively. In the
weak-coupling case $0 < X << t$, a standard BCS-type mean-field
approximation \cite{hir} and a renormalization-group analysis in
the one-dimensional (1D) continuum-limit theory \cite{jap}, show
that a small positive $X$ gives rise to an effective atractive
interaction for a particle density $ n > 1$, while this interaction
is repulsive for $n < 1$, and vanishes at half filling. This situation cannot
be extended to the case $X=t$, since for
these parameters ($t_{AB} = t_{AA} + t_{BB} = 0$), $H_t$ is
symmetric under an electron-hole transformation and the physics for
densities $n$ and $2-n$ should be the same. Thus, it is of interest
to study this case. This is one of the goals
of this Letter. Strack and Vollhardt studied the model for these
parameters (including $V$) at half filling and argued that this
case correspond to a physically relevant range of parameters \cite{s2}.
\par The study of the Mott transition also makes the case $t_{AB} =
0$ appealing, because of the supression of antiferromagnetic correlations.
This avoids the problem of having to distinguish between a
Mott insulator in which the particles become localized as a
consequence of strong on-site repulsion and an antiferromagnetic
insulator, in which a weak interaction opens a gap in a nested
Fermi surface. The latter is the case of the Hubbard model in
bipartite lattices. Studies of the Mott transition in these cases
are restricted to the paramagnetic phase \cite{bri,kot,roz}. Other
studies have taken nonbipartite lattices \cite{kris} or systems in
which the noninteracting Fermi surface has no nesting
\cite{so,shi}. In the large $U$ limit, the model of Eq.(1) becomes
equivalent to a generalized $t-J$ model \cite{cas} with hopping
$t_{AA} (t_{BB})$ for $n < 1 (n > 1)$, correlated hopping
$t^2_{AB}/U$, and antiferromagnetic exchange interaction
$J = 4t^2_{AB}/U$ which vanishes for $t_{AB} = 0$.
\par In this Letter we obtain the exact solution of Hamiltonian (1)
for a chain with open boundary conditions under the only
restriction $t_{AB} = \mid t_{AA} \mid - \mid t_{BB} \mid = 0$. We also discuss
the effect of a finite $t_{AB}$ on the basis
of our Lanczos results for finite chains. Strack and Vollhardt
obtained the exact ground state for $t_{BB} = - t_{AA} = t$, for arbitrary
dimension including the nearest-neighbor repulsion $V$, but only
for $n = 1$ and two regimes of parameters in which all particles
are static in the ground state \cite{s2}. In 1D and for $V = 0$ we are able
to obtain all eigenstates for arbitrary filling, particularly in a
third regime of parameters in which the dynamical part of the
Hamiltonian $H_t$ plays an important role in the ground state.
\par The exact solution of the model is greatly facilitated by its
symmetries. In any dimension for $t_{AB} = 0, [H_t,H_U] = 0$ and
the number of doubly occupied sites is conserved \cite{s2}. Also,
as in the case of the model of Essler, Korepin and Schoutens
\cite{es2}, for $t_{AB} = 0, H_t$ commutes not only with the total spin, but
also
with the following generators of another SU(2) algebra:
\begin{equation}
\eta = \sum^L_{i = 1} c_{i \downarrow} c_{i \uparrow} ,~ \eta^\dagger =
\sum_{i = 1}^L c^\dagger_{i \uparrow} c^\dagger_{i \downarrow} ,~ \eta_z =
\sum^L_{i = 1} ( \frac{1}{2} - \sum_\sigma c^\dagger_{i \sigma} c_{i
\sigma}) ,
\end{equation}
where $L$ is the number of sites. This allows us to construct
eigenstates of minimum energy which posess off-diagonal long-range
order for sufficiently small values of $U$ and $\mid n-1 \mid$.
\par The solution of the chain is obtained mapping $H_t$ into a
tight-binding model of spinless fermions. To obtain this mapping it
is convenient to write $H$ in a slave-boson representation. We
represent the four possible states at site $i: \mid 0 >,
c^\dagger_{i \sigma} \mid 0 >, c^\dagger_{i \uparrow} c^\dagger_{i \downarrow}
\mid 0>$, by $e^\dagger_i \mid 0 >, f^\dagger_{i \sigma} \mid 0 >,
d^\dagger_i \mid 0 >$ (pictorially $\circ, \uparrow$ or
$\downarrow$ and $\bullet$) respectively, using two bosons to
represent the empty $(\circ)$ and doubly occupied $(\bullet)$ sites
and two fermions ($\uparrow$ and $\downarrow$) to describe the
singly occupied sites. The Hamiltonian takes the form
\begin{eqnarray}
H = U \sum_i d^\dagger_i d_i + t_{AA} \sum_{{<ij>}_{\sigma}} f^\dagger_{j
\sigma} f_{i
\sigma} e^\dagger_i e_j \nonumber \\
- t_{BB} \sum_{{<ij>}_{\sigma}} f^\dagger_{j \sigma} f_{i \sigma} d^\dagger_i
d_j + 2
t_{AB} \sum_{<ij>} (f^\dagger_{j \uparrow} f^\dagger_{i \downarrow} e_i d_j
\nonumber \\
+ h.c.),
\end{eqnarray}
with the constraints $e^\dagger_i e_i + d^\dagger_id_i + \sum_{\sigma}
f^\dagger_{i
\sigma} f_{i \sigma} = 1$.
When $t_{AB} = 0$, the numbers $N_{\sigma} = \sum_i f^\dagger_{i \sigma}
f_{i \sigma}, N_e = \sum_i e^\dagger_i e_i$ and $N_d = \sum_i d^\dagger_i d_i$
are separately conserved. Note also that in a bipartite lattice,
changing the phase of the bosons $e_i$ or $d_i$ by (-1) in one
sublattice changes the sign of $t_{AA}$ or $t_{BB}$ respectively.
Thus we can choose these signs arbitrarily. Taking $-t_{AA} =
t_{BB} = t > 0$ as in Ref.\cite{s2}, $H_t$ takes the form:
\begin{equation}
H_t = -t \sum_{<ij> \sigma} [ f^\dagger_{j \sigma} f_{i \sigma} (e^\dagger_i
e_j
+ d^\dagger_i d_j) + h.c.]
\end{equation}
\par In a chain with open boundary conditions also the {\it order}
of the bosons and that of the fermions along the chain are
separately conserved: $H_t$ permutes the order of a fermion and a
boson which are nearest neighbors, but two bosons or two fermions
cannot be permuted. For a given number of fermions $N_f =
N_{\uparrow} + N_{\downarrow}$, let us numerate the $L$ sites,
$N_f$ fermions and $N_b = N_e + N_d = L - N_f$ bosons with similar
sequence (for example from left to right) using the labels $i,j$
and $m$ respectively. Then, any state with definite number of
particles on each site can be written as:
\begin{eqnarray}
\mid \psi_l > = \prod^{{N_b}}_{m = 1} [B(m)e^\dagger_{i(m)} + (1-B(m))
d^\dagger_{i(m)}] \nonumber \\
\times \prod^{N_{f}}_{j = 1} [ F(j) f^\dagger_{i(j) \uparrow} + (1-F(j))
f^\dagger_{i(j)
\downarrow}] \mid 0 > .
\end{eqnarray}
Here $i(m)$ is the position of the $m$th boson in the sequence
(its inverse, defined on the set of sites for which $n_{bi} = e^\dagger_i
e_i + b^\dagger_i b_i = 1$ is simply $m(i) = \sum^i_{l=1} n_{bl}$), and
$i(j)$ has a similar meaning for the fermions. $B(m) = 1$ if the
$m$th boson is an ``empty'' one and zero otherwise. Similarly in
terms of the spin of the fermions $F(j) = 1/2 + S^z_{i(j)}$. The
products are ordered throughout with increasing labels to the
right. As an example the state $\mid \psi_l > = \circ \uparrow
\downarrow \circ \bullet \uparrow \uparrow \downarrow \bullet$ ...
and any other state $\mid \psi_{l'} >$ such that $<\psi_l \mid H_t
\mid \psi_{l'} > \neq 0$ have $B(1) = B(2) = 1, B(3) = B(4) = 0,
F(1) = F(3) = F(4) = 1$ and $F(2) = F(5) = 0$.
\par Due to the properties of Eq.(4) and the open boundary
conditions, the 1D model has an extremely rich symmetry structure,
including $L$ SU(2) symmetries which are the local versions of
those previously mentioned.  There is one usual spin SU(2) algebra
related to each of the $N_f$ fermions and a ``local pairing'' SU(2)
algebra related with each boson. As an example it can be easily
verified that $(H_t e^\dagger_{i(m)} d_{i(m)} - e^\dagger_{i(m)} d_{i(m)} H_t)
\mid \psi_l > = 0$, where $e^\dagger_{i(m)} d_{i(m)}$ is a raising operator.
Thus one can separately diagonalize $H_t$ in each subspace of
definite values of $B(m)$ and $F(j)$. For fixed $N_f$ there are
$2^L$ subspaces and the size of each one is $( ^{L}_{N_f})$.
The raising and lowering operators establish a one to one
correspondence between each state of one of these subspaces and the
corresponding one of another subspace and $H_t$ takes the same form
in all these subspaces. In the subspace of highest weight of all SU(2) algebras
(all
$B(m) = F(m) = 1)$, the solution of $H_t$ for given $N_f$ is easily
obtained. The eigenstates, written in the original representation
have the form:
\begin{equation}
\mid \psi^0_e > = \prod^{{N_f}}_{j=1} c^\dagger_{{k_j}\uparrow} \mid 0> ,
{}~c^\dagger_{k \uparrow} = (\frac{2}{L+1})^{1/2} \sum_i \sin (ki) c^\dagger_{i
\uparrow},
\end{equation}
where the possible values of $k(L+1)/ \pi$ are positive integers.
These eigenstates can be extended to any values of $B(m)$ and
$F(j)$ using the lowering operators:
\begin{eqnarray}
\mid \psi_e > = \prod^L_{i=1} \left \{ n_{fi} [F(j_i) + (1-F(j_i)) c^\dagger_{i
\downarrow} c_{i \uparrow}] + \right. \nonumber \\
\left. + (1-n_{fi}) [B(m_i) + (1 - B(m_i)) c^\dagger_{i \uparrow} c^\dagger_{i
\downarrow}] \right \} \mid \psi^0_e >,
\end{eqnarray}
where $n_{fi} = n_i(2-n_i)$, $n_i = \sum_{\sigma} c^\dagger_{i \sigma} c_{i
\sigma},
j_i  = \sum^i_{L=1} n_{fi}$ and $m_i = i-j_i$.
\par Eqs. (6) and (7) also describe all the eigenstates of $H = H_t
+ H_U$. The latter term reduces the degeneracy to $2^{{N_f}}
(^{N_b}_{N_d} )$ and adds $UN_d$ to the energy.
\par For each particle density $n$, the ground state of $H$ is
obtained minimizing the energy with respect to the density of
doubly occupied sites $d = N_d/L$ and taking the lowest $N_f$
values of $k$ in Eq.(6), with the constraint $nL = 2dL + N_f$. The
result is very simple. In the thermodynamic limit three regimes can
be distinguished depending on the values of $U/t$ and the particle
density $n$. Also three regions of values of $U/t$ can
be separated (For $n = 1$ and $\mid U \mid > 8t$ the ground state
was obtained previously by Strack and Vollhardt \cite{s2}):\\
\noindent a) $U > 4t$. This region lies inside what we call regime I: for $n
\leq 1$ the
physics is the same as that of a spinless model. The ground state
expectation value $<H_U> = 0$ and:
\begin{equation}
d=0 ,~ e(n) = -\frac{2t}{\pi} \sin (n \pi),
\end{equation}
where $e(n)$ is the energy density. For $n \geq 1$, from
electron-hole symmetry $d = n-1, e(n) = U(n-1) + e(2-n)$. For $n =
1, < H_t > = < H_U > = 0$ and the system is an insulator with
energy gap $U - 4t$.\\
\noindent b) $U < - 4t$. This region coincides with regime II. Here
(for an even number of particles) all particles are paired, all
pairs are static $(< H_t > = 0)$ and:
\begin{equation}
d = n/2 ,~ e(n) = U n/2 .
\end{equation}
c) $-4t \leq U \leq 4t$. In this region there are two critical
densities $n_1$ and $n_2$ defined by: $n_i = (1/ \pi) \arccos
(-U/4t)$ and $n_1 \leq 1 \leq n_2 = 2-n_1$. For $n \leq n_1$ or $n
\geq n_2$ the physics corresponds to regime I and the ground state
and its energy was described above. Instead, for $n_1 < n < n_2$
the system is inside regime III. This regime is the only one in
which empty, single and double occupancy at any site is possible,
and the competition between $H_t$ and $H_U$ is apparent in the
ground state. The double occupancy and energy are given by:
\begin{equation}
d = \frac{n - n_1}{2} ,~ e(n) = Ud - \frac{1}{2 \pi} (16t^2 - U^2)^{1/2}
\end{equation}
\par In regimes II and III the system is at the borderline of phase
separation and also of superconductivity. Eigenstates with
off-diagonal long-range order (ODLRO) are part of the degenerate
ground state. To show this, let us take an eigenstate $\mid \psi_g
>$ of the form of Eq.(7), with $N_d$ doubly occupied sites, which
belongs to the ground state. The state $\mid \psi > = \eta^{{N_d}}
\mid \psi_g >$ with $\eta$ given by Eq.(2), is clearly different
from zero (it is obtained from $\mid \psi_g >$ putting all $B(m) =
1$ in Eq.(7)) and is also an eigenstate of $H_t$ with the same
eigenvalue as that of $\mid \psi_g >$. Also $\mid \psi >$ is a
highest-weight state of the $\eta$-pairing SU(2) algebra (Eq.(2)).
Similarly the state $\mid \psi_{{N_d}} > = (\eta^\dagger)^{{N_d}} \mid
\psi >$  is an eigenstate of $H_t$ with the same eigenvalue, and an
eigenstate of $H$ with the same energy as the original state $\mid
\psi_g >$. In Ref. \cite{es2}, it is shown that $\mid \psi_{{N_d}}
>$ in the thermodynamic limit ($L \rightarrow \infty$ with $d =
N_d/L$ constant) has ODLRO if $d \neq 0$ and $1+d-n = N_e/L \neq 0$.
\par The model has a metal-insulator transition at $U_c = 4t$. The
four-boson theory of Kotliar and Ruchenstein \cite{kot} in the
mean-field approximation gives $U_c = 16t/ \pi$ \cite{sim} in good
agreement with the exact value. The approximation also gives a
reasonably accurate $U_c$ for the infinite-dimensional Hubbard
model \cite{roz}.
\par The form of the Hamiltonian in the representation of Eq.(3)
suggests that addition of a small $t_{AB}$ such that it can be
treated in second-order perturbation theory, introduces
antiferrogmagnetic correlations between nearest-neighbor fermions
and allows the permutation of nearest-neighbor bosons $d$ and $e$,
increasing their mobility and favoring superconductivity. We have
solved numerically the model for $t_{BB} = -t_{AA} = 1, t_{AB} =
0.2 and L = 10$.
For $1/2 < n \leq 1$, the model exhibits phase separation for
$U > U_s$ with $U_s \sim 1$ for $n \sim 3/4$ and $U_s=0$ for $n=1$,
while for $U<U_s$ the system behaves
as a Tomonaga-Luttinger liquid (TLL) \cite{oga}. For $n < 1/2$ the
TLL behavior is observed for all values of $U$. Within the TLL regime,
the evaluation of the compressibility, the Drude weight and
the spin and charge velocities allowed us to derive
the correlation exponent $K_{\rho}$ \cite{oga}. The resulting values
indicate that the dominant correlations are the superconducting ones
for $1/2 < n < 1$ and the charge-charge ones for $n <1/2$.
\par In this Letter we have solved exactly a Hubbard chain including
bond-charge repulsion for a particular value of the latter. The
model displays a Mott transition at half filling and in two regimes
of parameters the ground state contains superconducting states.
Numerical results show that superconductivity is favored by a small
perturbation for not too large on-site Coulomb repulsion.
\par One of us (L.A.) is supported by the Consejo Nacional de
Investigaciones Cient\'{\i}ficas y T\'{e}cnicas (CONICET),
Argentina. A.A.A. is partially supported by CONICET.

\newpage
\end{document}